# Cross-Detection and Dual-Side Monitoring Schemes for FPGA-Based High-Accuracy and High-Precision Time-to-Digital Converters

Daehee Lee, *Member, IEEE,* Minseok Yi, *Member, IEEE,* Sun Il Kwon, *Senior Member, IEEE*

*Abstract*— This study presents a novel field-programmable gate array (FPGA)-based Time-to-Digital Converter (TDC) design suitable for high timing resolution applications, utilizing two new techniques. First, a cross-detection (CD) method is introduced that minimizes the occurrence of bubbles, which cause inaccuracy in the timing measurement of a TDC in thermometer codes, by altering the conventional sampling pattern, thereby yielding an average bin size half of its typical size. The second technique employs dual-side monitoring (DSM) of thermometer codes, including end-of-propagation (EOP) and start-of-propagation (SOP). Distinct from conventional TDCs, which focus solely on SOP thermometer codes, this technique utilizes EOP to calibrate SOP, simultaneously enhancing time resolution and the TDC's stability against changes in temperature and location. The proposed DSM scheme necessitates only an additional CARRY4 for capturing the EOP thermometer code, rendering it a resource-efficient solution. The CD-DSM TDC has been successfully implemented on a Virtex-7 Xilinx FPGA (a 28-nm process), with an average bin size of 6.1 ps and a root mean square of 3.8 ps. Compared to conventional TDCs, the CD-DSM TDC offers superior linearity. The successful measurement of ultra-high coincidence timing resolution (CTR) from two Cerenkov radiator integrated microchannel plate photomultiplier tubes (CRI-MCP-PMTs) was conducted with the CD-DSM TDCs for sub-100 ps timing measurements. A comparison with current-edge TDCs further highlights the superior performance of the CD-DSM TDCs.

## I. Introduction

TIME-TO-DIGITAL converters (TDCs), known for their precision, have found widespread applications in diverse fields such as laser range finding [1], [2], space science [3], ultrasonic flow measurement [4], particle and nuclear physics [5], [6], [7], [8], [9], and even in the medical field with positron emission tomography (PET) [10], [11], [12], [13]. The capability of TDCs to accurately capture signal arrival times can significantly elevate the research quality in these disciplines. For instance, in PET systems, precise time measurements can boost the signal-to-noise ratio of the reconstructed images, such as those in time-of-flight (TOF) PET detectors [14], [15], [16].

A TDC can be implemented either through application-specific integrated circuits (ASICs) or field-programmable gate arrays (FPGAs) [17], [18], [19]. Although ASIC-based TDCs have historically outperformed their FPGA counterparts due to their adjustable features for circuit performance, FPGA-based TDCs have made strides with advancing FPGA circuit technology, achieving a comparable time resolution of a few picoseconds [20], [21], [22], [23]. Furthermore, FPGAs offer benefits such as shorter time-to-market, lower implementation costs, and re-programmability, making them an attractive choice for scientific experiments that require circuit updates depending on purpose[1], [13].

A variety of schemes exist to implement FPGA-based TDCs. Among these, the pulse-shrinking method uses the discrepancies in rise and fall times to measure input time, despite its lengthy conversion time [24], [25]. Multiple clock-based TDCs, which measure time based on phase differences of out-phased clocks, are resource-effective but often yield poorer time resolutions than other architectures [24], [26], [27], [28]. Most FPGA-based TDCs employ a delay line (DL), directly converting time information by capturing the signal propagation through the DL synchronized with the clock signal [20], [23], [28], [29], [30], [31], [32]. The DL is mostly implemented by programmable logic blocks in the FPGA [18], [19].

Multi-time measurement schemes like the wave-union (WU) or multi-chain (MC) methods can further enhance time precision. However, the WU method has been found unsuitable for advanced technology due to its long dead time for time conversion and some technical issues [33], [34], [35]. Similarly, the MC method, which utilizes several identical DLs for each time conversion, consumes a significant amount of resources. The Vernier TDC method, taking advantage of the propagation time difference, achieves improved time resolution but requires substantial resources [36], [37], [38], [39], [40]. High resource demands are characteristic of these techniques, posing

Manuscript received XXX; revised XXX; accepted XXX.
This work was supported by the National Institutes of Health grant R01EB033536 *(Corresponding author: Sun Il Kwon)*
This work did not involve human subjects or animals in its research.

The authors are with the Department of Biomedical Engineering of the University of California Davis, Davis, 95616, CA, USA (e-mail: mnmlee@ucdavis.edu; msyi@ucdavis.edu; sunkwon@ucdavis.edu).
Color versions of one or more of the figures in this article are available online at http://ieeexplore.ieee.org

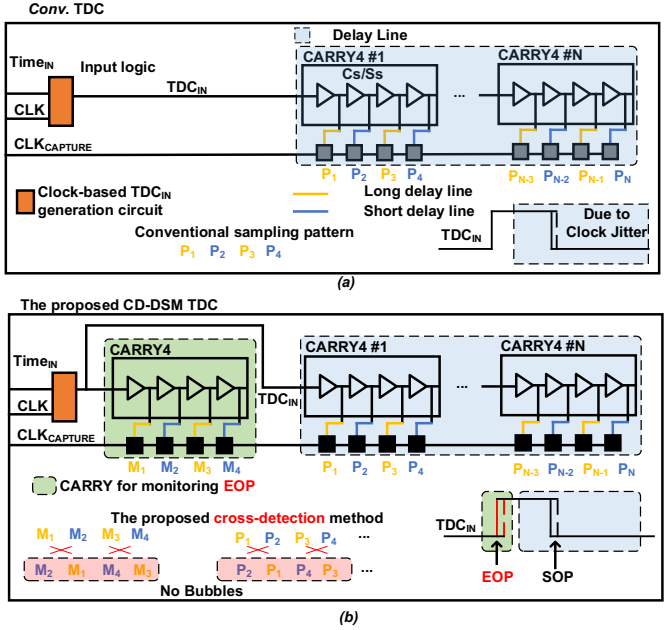

**Fig. 1.** Block diagram of a Conv. TDC (a) and the proposed Cross-detection & Dual-ended Monitoring (CD-DSM) TDC (b) which simultaneously captures not only SOP but also EOP signals.

challenges for TDC architecture design. However, if a design could achieve high time resolution while conserving resources, it would hold significant potential for multi-channel TDC applications, such as TOF PET detectors, where the objective is to implement as many channels as possible within limited FPGA resources [31], [41].

This study proposes two innovative schemes for a DL-based TDC to achieve high precision and high time resolution with low resource usage. Bubbles in thermometer code can cause incorrect digital output signals, reducing the precision of time measurements. These bubbles can arise due to signal propagation delays or metastable states within the FPGA TDCs [30], [42]. If not eliminated, they lead to erroneous time measurements, degrading the overall reliability and performance of the system. To address this, the first scheme applies the cross-detection (CD) method to significantly reduce bubbles in thermometer codes, thereby achieving improved time resolution. The second scheme monitors end-of-propagation (EOP) thermometer codes to calibrate conventional start-of-propagation (SOP) thermometer codes, which are influenced by process, voltage, and temperature (PVT) variations. This strategy not only enhances time resolution but also mitigates measurement errors in SOPs caused by changes in location and temperature. The performance of the proposed cross-detection and dual-sided monitoring (CD-DSM) TDC is subsequently compared with other studies.

## II. DESIGN

### A. Conventional DL TDC

In FPGAs, a key component for the effective implementation of digital circuits is the combinational logic block (CLB) [13], [31], [41]. These CLBs encompass lookup tables (LUTs) for executing combinational calculations, memory elements (like D flip-flops) for signal storage, and carry elements (such as the CARRY4 or CARRY8 in the Xilinx FPGA) for digital arithmetic. The conventional DL TDC, referred to as *Conv. TDC* in this paper, employs a carry chain that cascades CARRYs in sequence for signal propagation [13], [43].

A CARRY4 cell in Xilinx 7-series FPGAs has four carry outputs (Cs) and four sum outputs (Ss). However, connectivity constraints limit the linking of only four of the eight outputs to four D flip-flops within each CLB in the FPGAs [18].

The configuration of a *Conv. TDC* is depicted in Fig. 1(a). A time input ($Time_{IN}$) is initially processed by an input logic module, generating an input signal ($TDC_{IN}$) which is propagated through the DL. On the rising edge of a $CLK_{CAPTURE}$, the propagating $TDC_{IN}$ is captured at each D flip-flop for final time tagging [41], [43]. All reported DL TDCs focus solely on the SOP state for the final time conversion. However, it is worth noting that the carry chain used for signal propagation is vulnerable to clock jitter and PVT variations, both of which can introduce errors in the final time tagging [11], [18], [23], [44]. Since the SOP is captured by the clock ($CLK_{CAPTURE}$), it is highly sensitive to clock jitter, which can distort the measured time information.

### B. Cross-detection (CD) Method

D flip-flops capture thermometer codes, which are then converted into final time tags. However, due to internal delay discrepancies that are not in ascending order, bubbles frequently appear in acquired thermometer codes [33], [44]. If bubbles are present in the thermometer codes, the actual positions where the transitions should occur are not recognized, as the first transition position from the end of the thermometer code is used for the final time conversion. Therefore, the bubbles lead to a loss of time information, resulting in degraded time resolution. For instance, consider the scenario illustrated in Fig. 1(a), where the fourth propagating signal (P4) is captured by the fourth D flip-flop, which has a shorter time delay than the signal path for the preceding propagating signal (P3). As a result, P4 is consistently activated earlier than P3, leading to bubbles in P3 and causing the loss of time information between P4 and P3 due to the presence of these bubbles. This necessitates the use of bubble rejection circuits before converting thermometer codes to binary, thereby increasing resource usage [20], [41]. Others have proposed the use of a summation counter instead of bubble correction [45]. However, this approach still requires resources for the adder implementation while removing the bubbles.

We conducted an analysis of the bubble pattern and discovered that the even outputs of a CARRY4 (Virtex-7 Xilinx FPGA) are consistently triggered before the preceding outputs (e.g., P1-P2 and P3-P4). In response, we introduced the CD method, which switches each even output with the preceding odd output. However, it should be noted that the signal path from CARRY4 outputs to D flip-flops is sensitive to routing. Thus, we opted for sampling data by altering the thermometer code from the D flip-flops. For example, instead of the P1-P2-P3-P4 sequence, we acquired data in the P2-P1-P4-P3 sequence. The CD method is illustrated in Fig. 1(b). Remarkably, this technique requires neither additional resources nor manual tuning for bubble rejection, which often requires repeatable experiments. By removing the bubbles in thermometer codes, the time resolution is enhanced due to the reduction in bin size made possible by the recovery of the lost time information facilitated by the CD method.

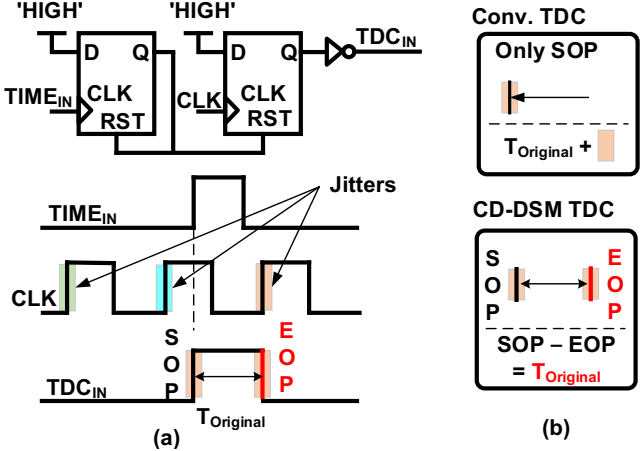

**Fig. 2.** Input logic to generate a $TDC_{IN}$ with SOP and EOP. The principle of DSM to acquire accurate time information compared to *Conv.*TDC.

*C. Cross-detection & Dual-ended Monitoring (CD-DSM) TDC*

The proposed CD-DSM TDC begins by employing an input logic module to generate a $TDC_{IN}$ signal that propagates through the following DL. As illustrated in Fig. 2(a), the $TDC_{IN}$ features two transitions for each $TIME_{IN}$: SOP and end-of-propagation (EOP). The input logic module initiates a 0-to-1 transition, referred to as SOP, whenever there is an incident $TIME_{IN}$. Subsequently, the EOP signal is generated to terminate the high state, causing a 1-to-0 transition synchronized with CLK, thereby preparing the TDC to capture the next $TIME_{IN}$.

In the *Conv.* TDC, only the SOP signal is captured using the $CLK_{CAPTURE}$ for time conversion. However, the clock used for capture is highly sensitive to variations in process, voltage, and temperature (PVT) conditions. Fluctuations in the clock's transition point can alter the SOP positions, leading to potential timing errors in SOP. As depicted in Fig. 2(b), clock jitter can make accurate measurement of the SOP challenging, resulting in a final time measurement that includes jitter information ($T_{Original}$ + jitter).

In this study, we propose a DSM technique that utilizes both SOP and EOP of each $TIME_{IN}$ for high-time resolution TDC, enabling resource-efficient implementation. The EOP, generated by the input logic and synchronized with the CLK, includes jitter. The EOP propagates through a single additional CARRY4 and is captured to reflect the propagation speed influenced by PVT conditions. By measuring both transitions using the same $CLK_{CAPTURE}$, the DSM TDC captures clock jitter and PVT-induced propagation errors in both EOP and SOP. The final time information corresponding to each $TIME_{IN}$ can be obtained by subtracting EOP from SOP. This subtraction cancels out the error components present in both signals, eliminating the need for additional error correction circuits and significantly reducing the required circuitry compared to *Conv.* TDCs with calibration circuits. This efficiency is particularly advantageous for multi-channel configurations.

Additionally, EOP thermometer codes, like SOP, also use a CARRY4 and are subject to bubbles in thermometer codes. The CD method, applied during final EOP signal sampling, significantly reduces these bubbles, as shown by the transformation from M1-2-3-4 to M2-1-4-3 patterns in Fig. 1(b). The proposed method, integrating both CD and DSM techniques, is referred to as CD-DSM TDC.

For optimal CD-DSM TDC operation, $CLK_{CAPTURE}$ is finely adjusted to locate EOP within a single CARRY4. In addition, since SOP and EOP exhibit different digital transition speeds for 0-to-1 and 1-to-0 transitions, respectively [13], a suitable correction factor should be determined to improve timing performance. Based on repeated experiments, a correction factor of 0.5 was found to be the most effective in achieving both the minimum bin size and the best linearity. For instance, when the difference in EOP is -1 compared to a reference EOP position, 0.5 of a thermometer code is added to the SOP. By employing this strategy, the time tag for each $TIME_{IN}$ can be accurately calculated.

III. EXPERIMENTAL SETUP

The CD-DSM TDC operated at a clock frequency of 550 MHz (CLK). To provide fine-range time conversion that slightly exceeds the clock period (1.818 ns), a DL was implemented using forty-three CARRY4s. In contrast, a 12-bit coarse counter was used for long-range time conversion [41]. The fine-counter value was obtained directly by capturing thermometer codes from the DL, and this value was combined with the coarse counter for the final time tag [2], [43]. Additionally, one CARRY4 was allocated for the EOP. In total, 176 thermometer codes (calculated from 43 CARRY4s × 4 outputs + 1 CARRY4 × 4 outputs) were conveyed to the computer through the UART, along with a 12-bit coarse counter value for each $TIME_{IN}$ tagging. The $CLK_{CAPTURE}$ captured the propagating signal on D flip-flops at the rising edge, operating at the same frequency as the CLK but with a phase delay that positioned the EOP within a single CARRY4.

To substantiate the effectiveness of the CD-DSM TDC, the *Conv.* TDC and the *Conv.* TDC (CD), which employs CD into *Conv.* TDC, were implemented at the same physical location. To measure implementation site variations in one chip, the three TDCs were implemented and evaluated at three different random locations (Slice X0, X22, and X102). Data acquisition was then executed independently for each of the three TDCs. A function generator (AFG3252, Tektronix, USA) served as the $TIME_{IN}$ source, generating random inputs at a frequency asynchronous to the CLK frequency. A total of 90,900 samples were used for each code density test.

The performances of the three TDCs were compared by calculating the average bin size, RMS value, differential non-linearity (DNL), integral non-linearity (INL), linearity error, and other relevant metrics derived from the code density tests. Additionally, the resilience of the proposed CD-DSM TDC against location and temperature variations was evaluated.

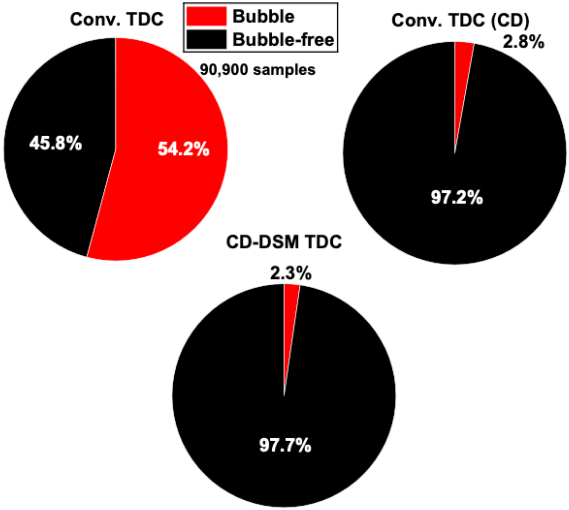

**Fig. 3.** Bubble ratio to the total sampled thermometer codes obtained from three different TDCs: *Conv.* TDC, *Conv.* TDC (CD), and CD-DSM TDC.

We conducted a series of tests to assess the performance in calculating the time difference between two identical CD-DSM TDC channels. Time differences were generated by splitting a signal from the function generator using a combination of three cables of different lengths: Cable A, B, and C (8 cm, 30.3 cm, and 50.2 cm, respectively). Each time measurement was then fitted to a Gaussian distribution. From these fits, we calculated the full width at half maximum (FWHM) and used the center value for linearity evaluation. The resulting data were subsequently compared to data obtained from an oscilloscope (MSO64B, Tektronix) with a bandwidth of 2.5 GHz and a 20 ps sampling time. Furthermore, we utilized the A-C cable setup to conduct a temperature-dependent performance test. Measurements were taken at various ambient temperatures, ranging from 0 to 50 °C in 10 °C increments.

## IV. RESULTS AND DISCUSSION

### A. Bubbles in Thermometer Codes

The prevalence of bubbles in the thermometer codes of the three different TDCs (*Conv.* TDC, *Conv.* TDC (CD), and CD-DSM TDC) was analyzed. The results, obtained from 90,900 samples for each TDC, are displayed in Fig. 3. The *Conv.* TDC showed that 54.2% of the total acquired samples contained bubbles. However, the use of the CD method in the *Conv.* TDC (CD) led to a significant decrease (97.2%) in the presence of bubbles for the acquired thermometer codes. There was a marginal increase to 97.7% in the proportion of bubble-free thermometer codes with the CD-DSM TDC. The presence of bubbles in EOP had a minor impact on the overall bubble ratio.

This improved performance was attributed to the proposed CD method, which only required a modification in the sampling sequence, eliminating the need for additional bubble correction circuits. Thus, the CD-DSM TDC achieved a significant reduction in bubbles, with only 2.3% of thermometer codes containing bubbles, without requiring an extra circuit for bubble correction. It is important to note that this marginal 2.3% of bubbles translated into less than a 1-picosecond difference when measured across 1,000 random data points. This small trade-off in accuracy was deemed acceptable for our design due to the elimination of a bubble correction circuit, providing a more efficient TDC system.

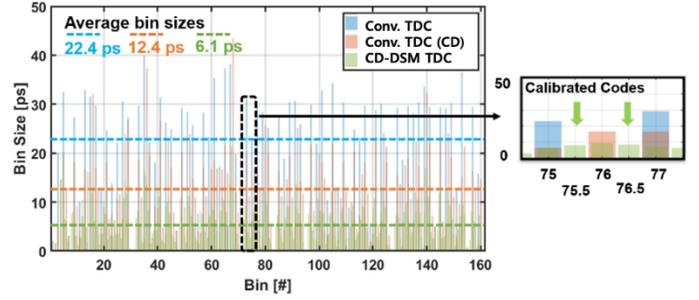

**Fig. 4.** Code density test results of three different TDCs: Conv. TDC, Conv. TDC (CD), and CD-DSM TDC. The insert box represents the magnified 75-77 bin range.

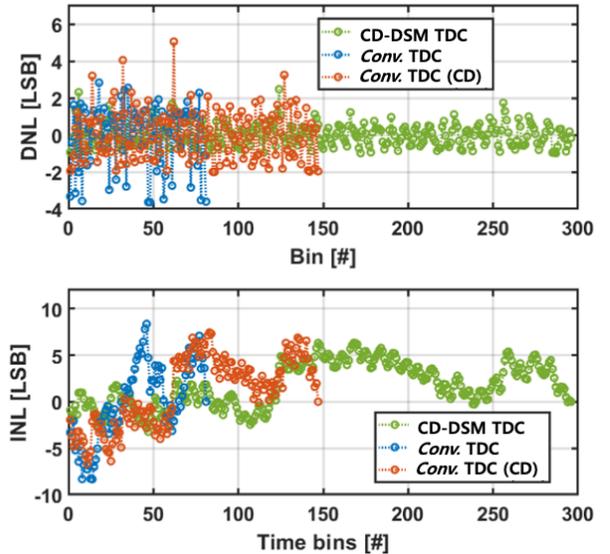

**Fig. 5.** DNLs (top) and INLs (bottom) for three different TDCs: *Conv.* TDC, *Conv.* TDC (CD), and CD-DSM TDC.

### B. Performances for Single-channel TDCs

Fig. 4 illustrates the time bin distribution from the code density tests for the three TDCs. The total number of bins for one CLK period (1.818 ns) was 81, 147, and 296 for the *Conv.* TDC, the *Conv.* TDC (CD), and the CD-DSM TDC, respectively. The *Conv.* TDC's average bin size was 22.4 ps with a standard deviation (SD) of 11.1 ps, and a single-shot accuracy (SD divided by $\sqrt{2}$ [21]) was 7.9 ps. These results align with other DL TDCs using identical 28-nm process FPGA chips [39]. The CD method notably reduces thermometer code bubbles and average time bin size, achieving a bin size of 12.4 ps with a single-shot accuracy of 5.8 ps. The CD-DSM TDC, by

TABLE I
MEASURED TDC PERFORMANCES AT THE THREE DIFFERENT IMPLEMENTATION SITES

| | Slice X0 | | | Slice X22 | | | Slice X102 | | |
|---|---|---|---|---|---|---|---|---|---|
| | Conv. TDC | Conv. TDC (CD) | CD-DSM TDC | Conv. TDC | Conv. TDC (CD) | CD-DSM TDC | Conv. TDC | Conv. TDC (CD) | CD-DSM TDC |
| Average bin [ps] | 22.4 | 12.5 | 6.1 | 22.4 | 12.5 | 6.1 | 22.4 | 12.4 | 6.1 |
| RMS | 10.8 | 8.2 | 4.5 | 10.8 | 8 | 3.9 | 10.8 | 8.3 | 3.8 |
| DNL [min max] LSB* | [-3.6 2.8] | [-2.0 2.9] | [-0.9 3.3] | [-3.6 3.0] | [-2.0 4.8] | [-0.9 2.5] | [-3.6 2.8] | [-2.0 5.1] | [-0.9 2.5] |
| INL [min max] LSB* | [-7.9 8.7] | [-9.0 5.7] | [-4.8 6.5] | [-8.3 7.0] | [-6.9 6.0] | [-2.7 7.9] | [-8.3 8.4] | [-6.7 7.4] | [-3.2 6.4] |
| R-square | 0.9991 | 0.9991 | 0.9994 | 0.9992 | 0.9994 | 0.9995 | 0.9987 | 0.9992 | 0.9995 |
| Linearity [min max] ps | [-29.8 62.5] | [-44.7 39.8] | [-24.8 42.5] | [-30.7 50.0] | [-30.9 42.1] | [-11.0 51.4] | [-32.3 61.2] | [-29.4 51.1] | [-15.7 41.9] |
| Linearity error [ps] | 23.1 | 24.4 | 16.4 | 20.4 | 18.6 | 15.2 | 21.6 | 20.6 | 13.8 |

* LSB represents the LSB of the CD-DSM TDC

correcting SOP with EOP, further improved the bin size to 6.1 ps and increased the single-shot accuracy by 53.4%, achieving 2.7 ps compared to the *Conv.* TDC (CD). Additionally, Fig. 4 displays a magnified view of the range from bins 75 to 77. Consistent with the correction methods outlined in *II. Design C*, applying a 0.5 factor to SOPs with EOPs positions them between integer bins, thereby reducing not only the average bin size but also the standard deviation of the bins.

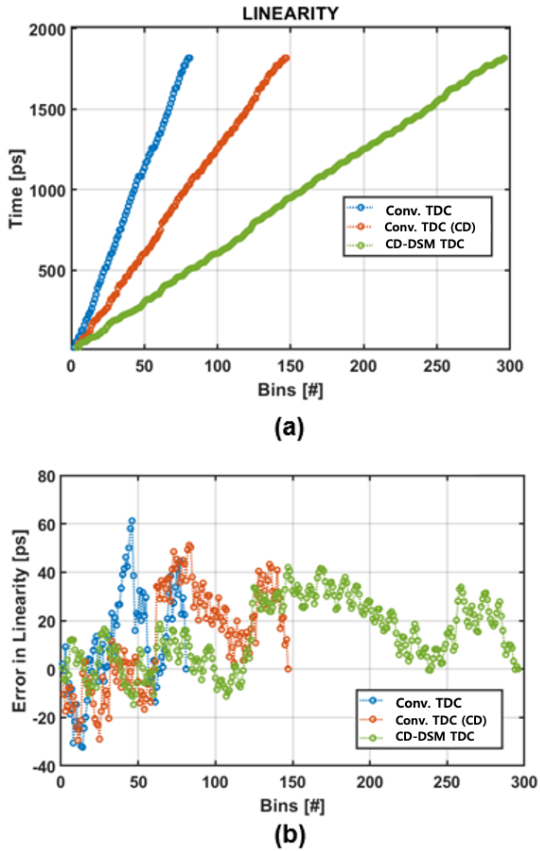

**Fig. 6.** Linearity of three TDCs (a) and the corresponding linearity errors for each line compared to the ideal linearity (b).

Fig. 5 presents the DNLs and INLs of the three TDCs. DNL and INL ranges, based on the CD-DSM TDC's least significant bit (LSB) of 6.1 ps, are compared in [Min Max] format. The CD-DSM TDC achieved a DNL range of [-0.9 2.5] LSB, a 36% reduction compared to the *Conv.* TDC. The INLs were [-8.3 7 8.4], [-6.7 7.4], and [-3.2 6.3] LSBs for the *Conv.* TDC, the *Conv.* TDC (CD), and the CD-DSM TDC, respectively. The INL full range in LSB also reduced from 15.7 LSB (*Conv.* TDC) and 14.1 LSB (*Conv.* TDC (CD)) to 9.5 LSB for the CD-DSM TDC.

In Fig. 6(a), the linearity of the TDCs over one CLK period is represented, characterized by a straight line from 0 to 1818 ps within the total bin range. As shown in Fig. 6(a), linearity is smoother with the CD-DSM TDC. Fig. 6(b) shows the errors compared to ideal linearity for each TDC. The SDs of linearity errors were 21.6, 20.6, and 13.8 ps, and the R-squared values were 0.9989, 0.9992, and 0.9995 for the *Conv.* TDC, the *Conv.* TDC (CD), and the CD-DSM TDC, respectively. These results demonstrate the superior linearity of the CD-DSM TDC.

Table I shows the results from the different TDCs, which were implemented at three different random locations (Slice X0, X22, and X102) to assess implementation site variations. Comparing the average bin size, the CD-DSM TDC consistently performed with a bin size of 6.1 ps across all three locations. Additionally, in terms of RMS, the CD-DSM TDC showed superior performance, with values ranging from 3.8 to 4.5 ps. The CD-DSM TDC exhibited DNL ranges from -0.9 to 3.3 LSB and INL ranges from -4.8 to 6.5 LSB across all tested locations. In summary, the results in Table I clearly demonstrate the CD-DSM TDC's enhanced performance in terms of time resolution, linearity, and non-linearity compensation across different locations.

TABLE II
MEASURED TIME DELAY PROPERTIES

| Temp. 24°C | TDC Type | Cable A-C (-Δ42.2) | Cable A-B (-Δ22.3) | Cable B-A (Δ42.2) | Cable C-A (Δ42.2) |
|---|---|---|---|---|---|
| Center of fitting [ps] | *Conv.* TDC | -1764.4 | -778.9 | 1169.9 | 2125.7 |
|  | *Conv.* TDC (CD) | -1763.5 | -779.5 | 1170.1 | 2125.8 |
|  | CD-DSM TDC | -1758.7 | -776.0 | 1170.2 | 2124.7 |
|  | Osc. | -1989.3 | -1072.5 | 1012.4 | 2000.4 |
| FWHM [ps] | *Conv.* TDC | 29.8 | 30.6 | 32.2 | 32.1 |
|  | *Conv.* TDC (CD) | 20.7 | 30.2 | 29.9 | 24.8 |
|  | CD-DSM TDC | 24.3 | 29.4 | 28.4 | 28.3 |
|  | Osc. | 27.4 | 33.6 | 23.1 | 34.2 |

*Values in parentheses indicate physical cable length differences in cm. The reference cable is Cable A.*

### C. Measuring Time difference using Two Identical TDCs

The evaluation of time difference measurements between two input signals was conducted using two identical TDCs. Three different-length cables were used to generate various time differences, and the results were acquired from four distinct cable combinations. All measured data were histogrammed and fitted with a Gaussian function to calculate timing measurement resolution values. Additionally, each time difference was measured with an oscilloscope for comparison. Table II summarizes the measured data. The *Conv.* TDC and *Conv.* TDC (CD) indicated average timing resolution values of 31.2 ps FWHM and 26.4 ps FWHM, respectively. The CD-DSM TDC exhibited 27.1 ps FWHM, while the oscilloscope data showed 29.6 ps. The CD-DSM TDC consistently provided stable timing resolution measurements, indicating values similar to those obtained from the oscilloscope. When comparing the linearity of the measured centroid values with the ideal values and the linearity determined through $R^2$ values, the CD-DSM TDC demonstrated the best linearity among the three TDCs.

The temperature dependency of each TDC was investigated using a cable A-C combination, which provides the largest time difference among the given cable combinations. Fig. 7 presents the measured time differences (center of the Gaussian fit). As the ambient temperature increased, the measured time difference between cables A and C decreased for both the *Conv.* TDC and the *Conv.* TDC (CD). On the other hand, the CD-DSM TDC exhibited more stable measurements regardless of temperature changes. This stability can be attributed to the DSM method, which compensates for the decreasing trend of SOP observed in conventional TDCs with EOP.

Fig. 8 demonstrates the distribution of EOP codes in response to changes in ambient temperature. Using the CD method, the thermometer code sequence for EOP codes was identified as M2-1-4-3. Temperature determined the location of the highest count, with M3, located at the end of the EOP sequence, showing the greatest number of counts among the four codes at 0 °C. However, as the temperature increased, the maximum count bins consistently transitioned towards lower thermometer code positions (M2), indicating slower propagation due to the decrease in EOP propagation speed. The prominence of M4 over a wide temperature range, as shown in Fig. 8, aligns with the reported fact that the time bin sizes of M2 and M4 in CARRY4 are wider than those of M1 and M3 [13]. Consequently, within the given ambient temperature range (10 °C – 30 °C), M4 exhibits dominance over a wider range. The changes in measured time with temperature, as shown in Fig. 7, used the EOP trend for SOP calibration, resulting in the CD-DSM TDC maintaining stable time measurement performance.

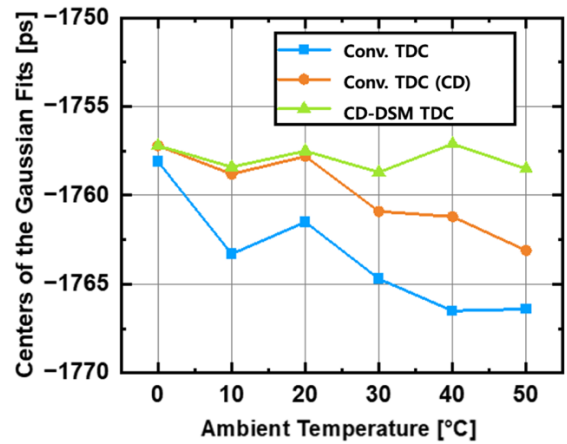

**Fig. 7.** Measured time differences with cable A-C recorded at six different ambient temperatures.

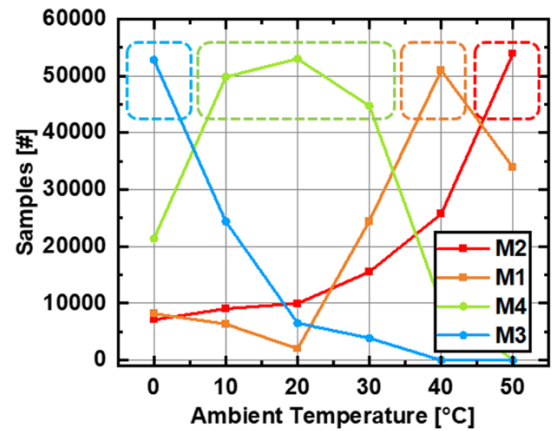

**Fig. 8.** Distribution of EOP codes as a function of ambient temperature with CD method applied

### D. Coincidence Timing Measurement with Two Ultrafast CRI-MCP-PMTs

The recent evolution in PET detectors has required the ability to measure time differences on the order of a few tens of

TABLE III
COMPARISON OF RECENT FPGA-BASED TDCS USING DLS

|  | **This work (DSM TDC)** | IEEE TNS 2017 [21]# | IEEE TIE 2018 [22] | IEEE TIE 2018 [22] | RSI 2019 [20] | RSI 2020 [48] |
|---|---|---|---|---|---|---|
| Average bin size [ps] | **6.1** | 1.15 | 10.5 | 5.02 | 5.8 | 3.3 |
| RMS [ps] | **3.8(single-shot) 5.4** | 3.5(single-shot) 4.9 | 14.6 | 7.8 | 6.6 | 5.4(single-shot) 7.6 |
| DNL[Min Max] LSB | **[-0.9 2.5]** | [-0.98 3.5] | [-1 3.78] Raw [-0.05 0.08]Ave | [-1 8.9] Raw [-0.12 0.11]Ave | [-0.98 3.25] | [-1 7.3] |
| INL [Min Max] LSB | **[-3.2 6.3]** | [-5.9 3.1] | [-0.88 5.90] [-0.09 0.11] | [-7.44 13.88] [-0.18 0.46] | [10.85 11.69] | [-3.2 20.8] |
| Resource usage | **22 LUTs* 176+22 FFs*** | 17169 LUTs | 1145 LUTs 1916 FFs | 704 LUTs 1195 FFs |  | 1603 LUTs 1178 FFs |
| Process | **28-nm** | 28-nm | 28-nm | 20-nm |  | 28-nm |

* Core implementation resources     # 20 channels of multi-DL

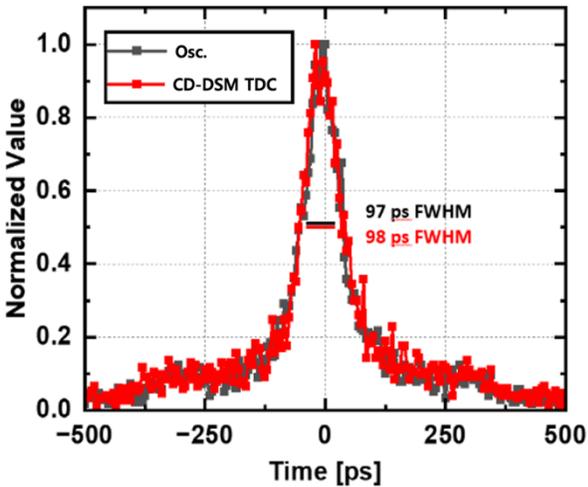

**Fig. 9.** Measured coincidence timing resolution (CTR) with the CD-DSM TDC and the oscilloscope

picoseconds [46]. To explore the feasibility of high-time resolution applications, two identical CD-DSM TDCs were fabricated, and their coincidence timing resolution (CTR) was calculated using two Cerenkov radiator integrated microchannel plate photomultiplier tubes (CRI-MCP-PMT) detectors and a $^{22}$Na radioactive point source. Small output signals from the two CRI-MCP-PMTs were amplified with additional amplifiers (ZFL-1000LN+, Mini-Circuits, bandwidth: 0.1 – 1000 MHz) for the oscilloscope and CD-DSM TDC inputs, simultaneously.

Fig. 9 presents the CTR results obtained using the CD-DSM TDCs and the oscilloscope with identical experimental setups. The limited bandwidth of the amplifiers introduced degradation to the original timing performance of the CRI-MCP-PMTs. Nevertheless, the CD-DSM TDC was capable of measuring CTR within a range of 98 ps FWHM, closely matching the 97 ps FWHM obtained from the oscilloscope. These results offer experimental confirmation of the high-time resolution capabilities of the CD-DSM TDC in practical applications.

*E. Comparison with Recent FPGA-based TDCs*

Table III draws a comparison between the proposed CD-DSM TDC and recently reported FPGA-based DL TDCs. Among the strategies listed, the TDC reported by Qin *et al.* [21] achieved superior performance in terms of average bin size, but at the expense of utilizing 20 multiple DLs to attain high-time resolution. However, excluding the TDC in [21], the proposed CD-DSM TDC exhibits a competitive average bin size while requiring fewer complex processes for high-time resolution achievement. In terms of overall RMS values, the CD-DSM TDC outperforms other TDCs. Upon evaluating the raw DNLs and INLs, the CD-DSM TDC presents relatively superior performance compared to other DL TDCs, excluding the TDC in [21], which required substantial resources.

The utilization of the proposed CD and DSM methods has proven effective in calibrating non-linearity in TDCs. It is noteworthy that the CD-DSM TDC was designed without peripheral circuits, such as bubble correction circuits, which could hinder a fair comparison of resource usage. Despite this, the core resource usage for the CD-DSM TDC is incredibly minimal, offering encouraging prospects for multi-channel applications such as TOF PET, which requires approximately a hundred channels per 5 cm × 5 cm area. The proposed methods enable the implementation of numerous TDCs in a confined area, delivering overall promising performance.

V. CONCLUSION

In this study, we introduced two novel techniques for high-time resolution in time-to-digital converters (TDCs) that require only a marginal increase in resources. The cross-detection (CD) method reduces the number of bubbles in thermometer codes by rearranging bins in a cross-sampling order. This is achieved by alternating the switching of even and odd sampling positions for the CARRY4's outputs, significantly enhancing time resolution without the need for additional logic implementation. The dual-side monitoring (DSM) method appropriately corrects start-of-propagation (SOP) signals with an end-of-propagation (EOP) position, enhancing the robustness of the final thermometer code in terms of time measurement performance under temperature and location variations.

The proposed CD-DSM TDC was implemented on a Virtex-7 board from Xilinx, utilizing a 28-nm process. The average bin size was reduced to a quarter of the original size found in conventional DL TDCs, which have almost the same resource use, from approximately 22 ps to 6 ps. The CD-DSM TDC also achieved a single-shot accuracy of 2.7 ps, which is a challenging value to attain even with finer processes such as the 20-nm process [22], [45], [47]. Additionally, the DSM methods led to reductions in both DNL and INL.

The time measurement performance exhibited stability between two identical CD-DSM TDCs, unaffected by fluctuations in location or temperature. The CD-DSM TDC demonstrated its efficacy by accurately measuring the ultra-narrow coincidence timing performance derived from two ultrafast radiation detectors, CRI-MCP-PMTs. Validated as a resource-efficient TDC, the CD-DSM TDC proves apt for multi-channel applications, including positron emission tomography (PET) systems, achieving successful measurement of sub-100 ps coincidence timing performance in a practical configuration.

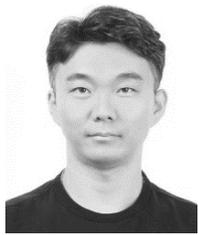

**Daehee Lee** (Member, IEEE) received the B.S. degree in electrical engineering from the Kyungpook National University (KNU), Daegu, South Korea, in 2010, and the M.S and Ph.D. degrees in nuclear and engineering from the Korea Advanced Institute of Science and Technology (KAIST), Daejeon, South Korea, in 2017. From 2017 to 2021, he was a senior researcher at the Agency for Defense Development (ADD), Daejeon, South Korea. S From 2021 to 2023, he was a Postdoctoral Scholar in the Department of Biomedical Engineering, University of California, Davis, CA, USA. Since 2023, he has been working as a scientist in the same department at the University of California, Davis, focusing on research areas that include novel gamma-ray detector development and system design for medical imaging technologies, especially time-of-flight positron emission tomography.

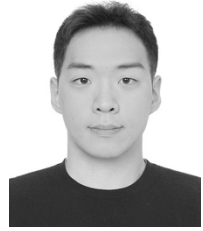

**Minseok Yi** received the B.S. (cum laude) degree in electrical and computer engineering from Seoul National University (SNU), Seoul, South Korea, in 2021. He is currently working toward the Ph.D. degree with the interdisciplinary program in bioengineering at SNU, since 2021. His current research interests include the development of novel gamma-ray detector development and system design for medical imaging technologies, especially time-of-flight positron emission tomography. Mr. Yi was the recipient of the BK21 Graduate School Innovation Project Group Colloquium Excellent Graduate School Student Award and the Conference Trainee Grant at the IEEE Nuclear Science Symposium and Medical Imaging Conference.

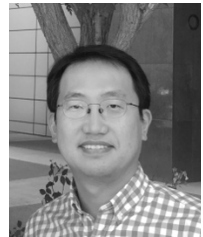

**Sun Il Kwon** (Senior Member, IEEE) received the B.S. degree in electrical engineering from the Korea Advanced Institute of Science and Technology (KAIST), Daejeon, South Korea, in 2002, and the Ph.D. degree in the interdisciplinary program of radiation applied life science from Seoul National University, Seoul, South Korea, in 2013. He is currently an Assistant Professor with the Department of Biomedical Engineering, University of California, Davis, Davis. His research interests include novel gamma-ray detector development and system design for medical imaging technologies, especially positron emission tomography.